\documentclass[pdflatex,sn-nature]{sn-jnl}

\usepackage{graphicx}%
\usepackage{multirow}%
\usepackage{amsmath,amssymb,amsfonts}%
\usepackage{amsthm}%
\usepackage{mathrsfs}%
\usepackage[title]{appendix}%
\usepackage{xcolor}%
\usepackage{textcomp}%
\usepackage{manyfoot}%
\usepackage{booktabs}%
\usepackage{algorithm}%
\usepackage{algorithmicx}%
\usepackage{algpseudocode}%
\usepackage{listings}%

\usepackage{graphicx}
\usepackage{upgreek} 

\usepackage{url}

\begin{document}

\title[]{Non-Hermitian Synthetic Phase Shifter: Topologically-Protected Phase Control via Tunable Losses }

\author[]{\fnm{Kevin} \sur{Zelaya}}
\equalcont{These authors contributed equally to this work.}

\author[]{\fnm{Jonathan} \sur{Friedman}}
\equalcont{These authors contributed equally to this work.}

\author[]{\fnm{Hector} \sur{Rubio}}

\author[]{\fnm{Stefan} \sur{Preble}}

\author*[]{\fnm{Mohammad-Ali} \sur{Miri}${}^*$}\email{ali.miri@rit.edu}

\affil[]{\orgdiv{Electrical and Microelectronic Engineering}, \orgname{Rochester Institute of Technology}, \orgaddress{\street{1 Lomb Memorial Drive}, \city{Rochester}, \postcode{14623}, \state{New York}, \country{USA}}}


\abstract{Phase shifters are fundamental reconfigurable components in photonic circuits. In conjunction with passive elements, they control light flow and serve as foundational building blocks for diverse applications, including communication, sensing, analog signal processing, and quantum control. Conventional phase shifters achieve phase control by modulating the refractive index through various physical mechanisms such as thermo-optic or electro-optic effects. However, despite expectations that such index‑based approaches would integrate seamlessly, they, in practice, restrict circuit size, bandwidth, and scalability and thus become bottlenecks to large‑scale photonic integration. Here, we introduce an alternative phase‑control approach based on optical loss modulation. We demonstrate a synthetic phase shifter that uses two independently controlled loss‑modulation stages combined with multipath interference to achieve full‑cycle phase tunability while maintaining constant amplitude. We develop a theoretical framework based on conserved topological charges to demonstrate how synthetic phase control can be achieved via non-Hermitian effects, enabling robust topologically-protected phase control. By shifting the paradigm from index control to loss modulation, the proposed synthetic phase shifter could pave the way for scalable integrated photonic systems that support applications from communications and sensing to photonic classical and quantum information processing.}


\maketitle

\section*{Introduction}
In photonics, phase control is typically achieved by modulating the refractive index of the dielectric materials via physical processes like the thermo-optic~\cite{liu2022thermo,harris2014efficient}, electro-optic~\cite{kang2019silicon,xin2011high,dushaq2024electro}, or acousto-optic effects~\cite{li2005optical}. Both the real (phase response) and imaginary (absorption and gain) parts of the refractive index can be manipulated. However, the ease of tuning of each component depends on the physical mechanism. Historically, the pursuit of power conservation (unitarity) for propagating fields has centered research on controlling the real part of the refractive index. This focus has yielded a plethora of programmable power-conserving optical computing units that leverage the spatial degrees of freedom~\cite{miller2013self,clements2016optimal,zelaya2024goldilocks,zelaya2025integrated}, and the intrinsic time and frequency domains of electromagnetic waves~\cite{piao2024programmable,friedman2025programmable}. In turn, there are scenarios where controlling absorption is more straightforward than precisely tuning the real part. Although this approach breaks the power-conservation paradigm, it enables relevant non-Hermitian effects~\cite{feng2017non,miri2019exceptional,deng2021parity}, induces richer non-unitary linear optical transforms~\cite{miller2013self,markowitz2024learning}, and facilitates devices with smaller footprints and higher speeds than conventional thermo-optic solutions~\cite{dave2024clearing}. 

Research exploring the use of amplitude-only control to achieve optical computing functionalities has not been widely exploited. This paper aims to fill this gap by introducing the theory, design, and characterization of a synthetic phase-control unit tuned exclusively via amplitude modulators (AMs). This AM-based approach is particularly compelling because amplitude modulators are typically smaller and faster than their conventional thermo-optic counterparts~\cite{chen2016study}, thereby enabling scalable and high-speed programmable architectures in photonics integrated circuits (PICs). Counterintuitively, we demonstrate that our proposed synthetic phase-control unit can replicate the functionality of a phase shifter by employing AMs as active elements in combination with waveguide arrays (WGA) as a wave-interference mechanism. This unit can perform a full, steady-state $2\pi$ phase rotation while maintaining a constant output amplitude, effectively synthesizing a phase-control operation from inherently lossy components. Moreover, the same device can be reconfigured to provide a steady phase response while enabling control of the output amplitude. The general theory developed is based on topologically protected quantities; thus, it is not constrained to a single implementation, is robust against defects, and the proposed solution can be adapted to different physical AM solutions, highlighting the versatility of our design.

\section*{Results}
\subsection*{Full Phase Control via Topological Protected Charges}
Current optical units that perform linear transforms on their input electric fields are built from basic building blocks that, when combined, enable richer linear operations. Phase-control components are among the most widely used solutions for actively controlling light flow in photonic circuits. Such phase-control units transform an input field $E$ into $Ee^{i\phi}$, with $\phi\in(0,2\pi)$ (left panel in Fig.~\ref{fig:Fig1}\textbf{a}). The added phase is insensitive to power measurements but plays a fundamental role in interferometry, such as in Mach-Zehnder interferometers (MZI). In turn, amplitude control is typically achieved in the lossy regime, transforming any incoming field to $\gamma E$, where $\gamma\in(0,1)$ (middle panel in Fig.~\ref{fig:Fig1}\textbf{a}). Indeed, photonic components like micro-ring and micro-disk resonators~\cite{bogaerts2012silicon,borselli2005beyond} can operate in the gain regime ($\gamma>1$), but they unavoidably induce resonances that may be undesirable when operating in broadband devices. Finally, electric field interference can be achieved using multiport couplers (left panel in Fig.~\ref{fig:Fig1}\textbf{a}), which perform linear operations on the input fields. Examples of this include waveguide arrays~\cite{Chr03,Wei16} and multimode interference (MMI) couplers~\cite{feng2008compact}.

The compact footprint and high-speed operability of AMs can be leveraged to induce phase-control functionalities. This is possible due to the complex-valued nature of the electric fields and interference effects. To illustrate the latter, let us consider two amplitude modulators $\gamma_{1,2}\in(0,1)$ and a complex-valued function, $z(\gamma_1,\gamma_2):=z_{\textnormal{R}}(\gamma_{1},\gamma_{2})+iz_{\textnormal{I}}(\gamma_{1},\gamma_{2})$, with $z_{\textnormal{R}}(\gamma_{1},\gamma_{2})$ and $z_{\textnormal{I}}(\gamma_{1},\gamma_{2})$ the real and imaginary parts of $z(\gamma_{1},\gamma_{2})$, respectively. The function $z(\gamma_{1},\gamma_{2})$ is then parameterized by the two real variables, and we thus explore the possibility of performing full phase control by tuning $\gamma_{1,2}$ alone. The exact form of $z(\gamma_1,\gamma_2)$ depends on the actual composition of the photonic device (Fig.~\ref{fig:Fig1}\textbf{b}), which can be built using combinations of passive multiport couplers and optical path deviations, in combination with two active AMs.

\begin{figure}[t!]
    \centering
    \includegraphics[width=0.9\linewidth]{Figures/PS_F1.png}
    \caption{\textbf{Lossy Phase-Control Unit.} \textbf{a} Building blocks constituting linear optical devices in photonics. The main components to manipulate the propagating electric fields include phase shifters for phase control, amplitude modulators for intensity attenuation, and N-port interferometers $\mathcal{U}\in U(N)$ for discrete linear interference. \textbf{b} Black-box device rendering a transfer function $z(\gamma_{1},\gamma_{2})$. This comprises only multiport couplers and amplitude modulators. \textbf{c}-\textbf{d} Particular realizations for phase-control units based on AMs using two-port 3 dB couplers (\textbf{c}) and three-port homogeneous WGAs (\textbf{d}). \textbf{e} Location of transmission zeros in the $\gamma_{1}-\gamma_{2}$ plane for different optical path deviations in the two-port (\textbf{c}) and three-port (\textbf{d}) cases. \textbf{f}-\textbf{g} Typical behavior of the transfer function $z(\gamma_{1},\gamma_{2})$ around an isolated zero. We illustrate the intensity (\textbf{f}) and phase ($\textbf{g}$) of $z(\gamma_1,\gamma_2)$, together with several contours surrounding the isolated transmission zero.}
    \label{fig:Fig1}
\end{figure}

To study the general requirements of the transfer function representing the device in Fig.~\ref{fig:Fig1}\textbf{b}, we consider $z(\gamma_{1},\gamma_{2})$ as a continuously differentiable function with respect to $\gamma_{1,2}$. We further assume it contains an isolated zero at some point $\Gamma_{0}\in\Gamma=\{(\gamma_{1},\gamma_{2})\vert \gamma_{1,2}\in(0,1)\}$. Then, suppose there exists a simple closed (non-self-intersecting) contour $\mathcal{C}\subset\Gamma$ encircling $\Gamma_{0}$. In that case, we can ensure that $\angle z\vert_{\mathcal{C}}$ spans a full $2\pi$ rotation when the parameters circulate the closed contour. This follows from Cauchy's argument principle~\cite{brown2009complex}, and it is equivalent to saying that the total accumulated phase is $2\pi \mathcal{W}$, with $\vert \mathcal{W}\vert$ a non-null integer denoting the winding number or topological charge. See Supplementary Information~\textcolor{red}{S1}. This is a fundamental result in topological insulators, widely used to quantify topological invariants, such as Chern numbers~\cite{bernevig2013topological,asboth2016short}, and exploited for topological light guiding in 1D photonic lattices~\cite{blanco2016topological}. Such a topological property makes the zero stable. That is, the phase is locked into the winding pattern around the trajectory $\mathcal{C}$, and a small and smooth change to the function cannot remove the zero. 

To visualize the intended operation, we consider two different relations for the phase-control unit based on two-port and three-port structures, as shown in Figs.~\ref{fig:Fig1}\textbf{c}-\textbf{d}, respectively. For the two-port case, we consider a construction using passive 3 dB couplers sequentially layered with two AMs. These AMs are positioned on different layers, as depicted in Fig.~\ref{fig:Fig1}\textbf{c}. If the AMs were situated on the same layer, they would produce a single effective amplitude parameter since a global parameter can always be factored out. In this case, it is straightforward to find that $z(\gamma_{1},\gamma_{2})=\gamma_{2}-\gamma_{1}-\gamma_{1}\gamma_{2}-1$, corresponding to the point `A' in Fig.~\ref{fig:Fig1}\textbf{e}, and an isolated transmission zeros do not exist in the loss modulation regime $\gamma_{1,2}\in(0,1)$. 

For the three-port device, we consider two unitary devices $\mathcal{U}_{1,2}$ interleaved with an AM layer, as shown in Fig.~\ref{fig:Fig1}\textbf{d}. Such passive unitaries can be implemented using MMIs or WGAs; in this work, we focus on the latter. In both configurations, we excite the uppermost input port and collect the output from the uppermost output port. We should thus chose $U_{1,2}$ so that the transmission zero lies in the domain $\gamma_{1,2}\in(0,1)$. Fig.~\ref{fig:Fig1}\textbf{e} depicts potential locations of the transmission zero, highlighting several scenarios. For instance, points `A' and `B' show zeros at the edge of the domain, leading only to open contours, and no full phase-sweep is covered in these cases. In contrast, point `C' indicates that the zero is outside the tunable domain for the AMs. Points `D' and `E' represent cases that achieve the desired placement of the transmission zeros. 

We are particularly interested in contours that maintain constant intensity and encircle the transmission zero. In such a case, the phase sweeps a full rotation ($\vert \mathcal{W}\vert=1$), enabling phase control with steady power via active amplitude modulators exclusively. Such a behavior is illustrated in Figs.~\ref{fig:Fig1}\textbf{f}-\textbf{g}. In this regard, the primary goal of this work is to design an equivalent photonic device capable of such a task. Notably, the complex-valued function $z(\gamma_{1}, \gamma_{2})$ is not tied to a specific form and can take various forms, allowing us to explore multiple tunable devices to achieve the desired functionality. For this purpose, we consider lossy (non-Hermitian) photonic units achieved via tunable AM elements, $\gamma_1$ and $\gamma_2$, positioned between multiple unitary multiport couplers. 

\subsection*{Lossy Photonic Integrated Circuit Design For Phase Control}
For this work, we devise a lossy PIC with tunable losses $\gamma_{1,2}$ so that its transfer function creates contours $\mathcal{C}$ in the $\gamma_{1}\textendash\gamma_{2}$ plane, examples of which are shown in Fig.~\ref{fig:Fig1}\textbf{e}, that enable control of the amplitude $(A(\gamma_{1},\gamma_{2}))$ and phase $(\phi(\gamma_{1},\gamma_{2}))$ of the output optical signal. In particular, it is of interest to find a trajectory that maintains constant amplitude while the phase evolves continuously from $0$ to $2\pi$. As noted earlier, there is no unique transfer function that achieves this goal. Thus, we aim for a minimal-component design that reduces the number of multiport couplers while keeping the number of tunable components fixed at two. To evaluate the proposed concept, we present a simple proof-of-concept device on a silicon-on-insulator (SOI) platform, tailored for open-access foundries.
We selected MZIs with integrated thermo-optic phase shifters to emulate amplitude modulation. Although other dedicated integrated solutions exist, such as phase-change materials (PCMs)~\cite{rios2022ultra,huang2025optical,hatayama2024phase} and SiGe-based modulators~\cite{douix2019sige}, MZIs offer a well-known performance and do not require specialized fabrication processes, making them ideal for our target platform. 

The layout design for the proposed phase-control unit is developed by employing two three-port waveguide arrays (passive layer) comprising homogeneous couplings interleaved with a layer of amplitude modulators (active lossy layer). For testing purposes, we incorporated three MZIs in the active layer. Nevertheless, we tune only two of the MZIs, leaving the third as a redundancy element. This design yields a three-port unit with a non-unitary transmission matrix 
\begin{equation}
M(\gamma_{1},\gamma_{2}):=F\Lambda F, \quad \Lambda:=\textnormal{diag}(\gamma_{1},\gamma_{2},1),
\label{eq:DUT}
\end{equation}
with $M(\gamma_{1},\gamma_{2})\in\mathbb{C}^{3\times 3}$ and $F$ the transfer matrix of a passive multiport coupler. In the current design, we utilize waveguide arrays as the multiport coupler $F$. To make the design more realistic, we incorporate additional passive effects in $ F$, such as the phases accumulated from uncompensated optical path deviations induced by fan-out at the ports. See Supplementary Information~\textcolor{red}{S2} for details. Despite its multiport nature, we operate the device by exciting the $n$-th input while measuring the intensity and phase of the $m$-th output, for $n,m\in\{1,2,3\}$. We thus recover the intended phase-control transfer function from the matrix component $M_{m,n}$ through the relation $z(\gamma_{1},\gamma_{2})=M_{m,n}(\gamma_{1},\gamma_{2})$.

\begin{figure}[t!]
    \centering
    \includegraphics[width=0.85\linewidth]{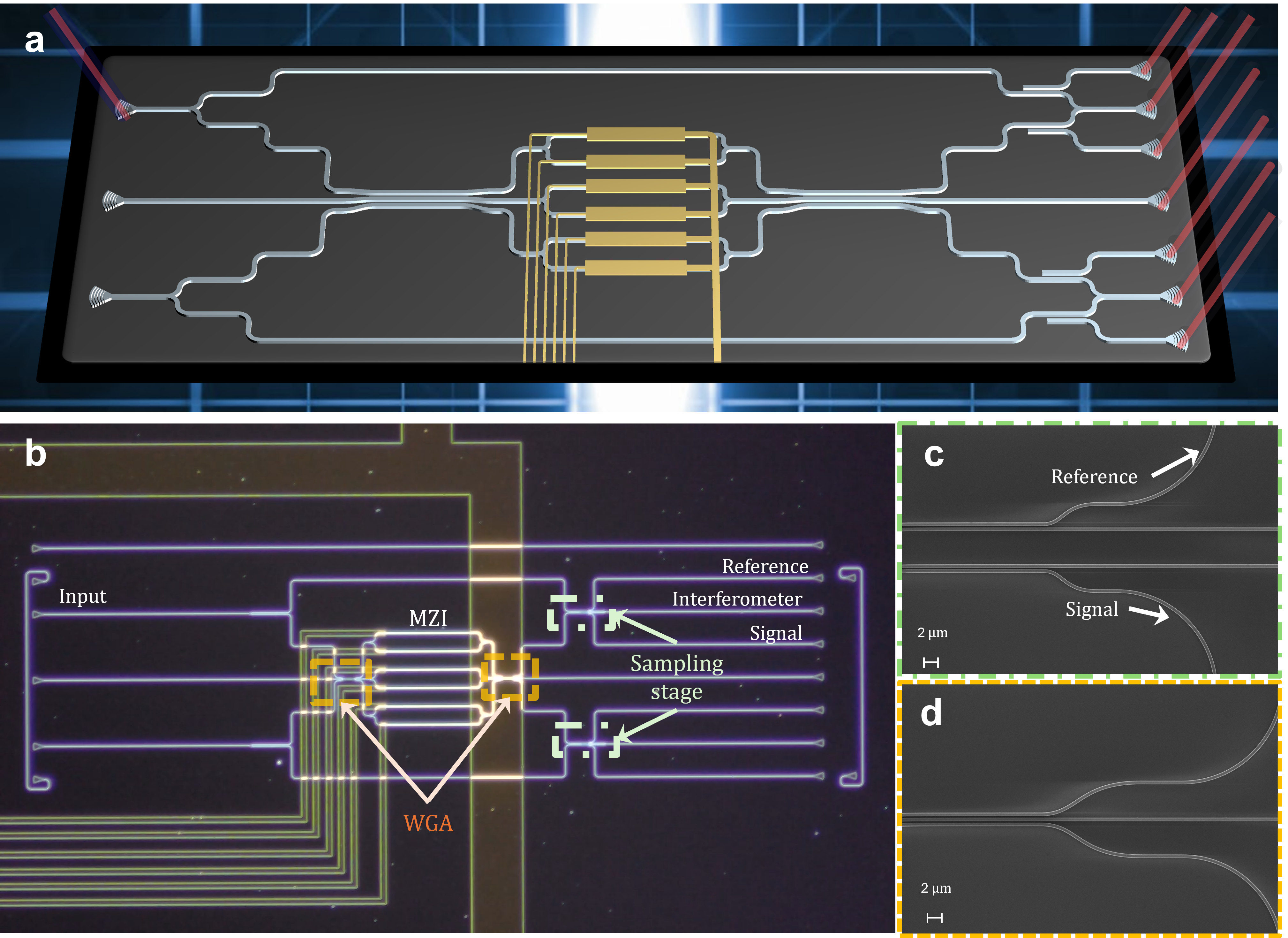}
    \caption{\textbf{Designed and Fabricated Phase-Control Unit.} \textbf{a} Render of the proposed PIC showcasing the DUT related to the phase-control unit, showcasing the interferometer stage used for phase retrieval, and the MZIs used as the AM elements. 
    \textbf{b} Microscope image of the fabricated sample. The blue and yellow glowing lines represent the optical waveguides and electrical wires, respectively. \textbf{c}-\textbf{d} Scanning electron microscope (SEM) image of the areas highlighted in the microscope image. This includes the SEM image of the sampling stage at the end of the interferometer (\textbf{c}) and the waveguide arrays (\textbf{d}). Both images show the routing waveguides that interconnect the PIC elements with the rest of the PIC.
    }
    \label{fig:Fig2}
\end{figure}

The device under test (DUT), depicted in Fig.~\ref{fig:Fig2}\textbf{a}, features a proposed phase-control structure and two embedded, symmetrical interferometers for phase retrieval. The phase-control unit consists of two WGAs, each with identical and homogeneously spaced waveguides interleaved with a layer of three MZIs for amplitude control. The uppermost interferometer splits the input signal into a reference arm, also referred to as the local oscillator (LO1), and a path to the DUT. The signal from the uppermost output port of the DUT is then combined with the LO1 arm to generate interference. A second, symmetrical interference process is implemented using the lowest ports, where a second independent local oscillator (LO2) is used. In each interferometer, a fraction of the power is sampled from the local oscillator and DUT signal ports via fixed-ratio directional couplers. This dual-interferometer setup enables operation across different port configurations and also serves as a redundancy protocol.

The photonic chip was fabricated using an open-access silicon foundry. The design features fully etched silicon (Si) waveguides placed on a 2 µm-thick SiO${}_{2}$ substrate. The silicon layer is fully etched, with a uniform thickness of 220 nm and a consistent width of 500 nm, maintained throughout this design. Additionally, the entire structure is covered with a 2.2 µm-thick SiO${}_{2}$ layer. For the active components, there are two metallization layers applied on top of the cladding: one for the microheater, which uses a high-resistance Ti/W alloy, and the other serves as the routing layer. Probing pads connect to the routing layer within the design area and are wire-bonded to a printed circuit board (PCB) to facilitate external electrical connectivity for each phase shifter.

Figure~\ref{fig:Fig2}\textbf{b} displays a microscope image of the photonic circuit. The white-glowing patterns indicate the fabricated waveguides, while the yellow-glowing areas illustrate the wire routing traces. Scanning electron microscope (SEM) images are also provided to showcase the fabrication details. Particularly, the SEMs of the WGAs (Fig.~\ref{fig:Fig2}\textbf{c}) and the sampling stage incorporating the fixed-ratio directional couplers (Fig.~\ref{fig:Fig2}\textbf{d}).

In this design, the WGAs interconnect with other elements in the circuit through cubic B\'ezier bends, which are designed to minimize crosstalk between neighboring waveguides while reducing the power loss due to the bend curvature. In turn, we incorporated circular bends with a 10 µm radius to fan out the WGA output and route light to the MZIs. As mentioned above, MZIs are used as a proof-of-concept solution to induce amplitude modulation in the circuit. Each MZI utilizes two Y-branches~\cite{zhang2013compact}, one for splitting of the incoming signal and the other for interference. Although two phase shifters are included in each MZI, only one is operated to induce the desired amplitude modulation.

The chip was packaged with electrical connections. This setup provides direct electrical access to the metal heaters for programming through an external source measurement unit (SMU), which is controlled via a computer interface. Two V-Groove assembly fiber arrays with a 127 µm pitch and an 8-degree angle were mounted on both ends of the chip to couple light between the input and output grating couplers and the relevant measurement equipment. 

\subsection*{Experimental Setup and Characterization}
\begin{figure}[t!]
    \centering
    \includegraphics[width=0.95\linewidth]{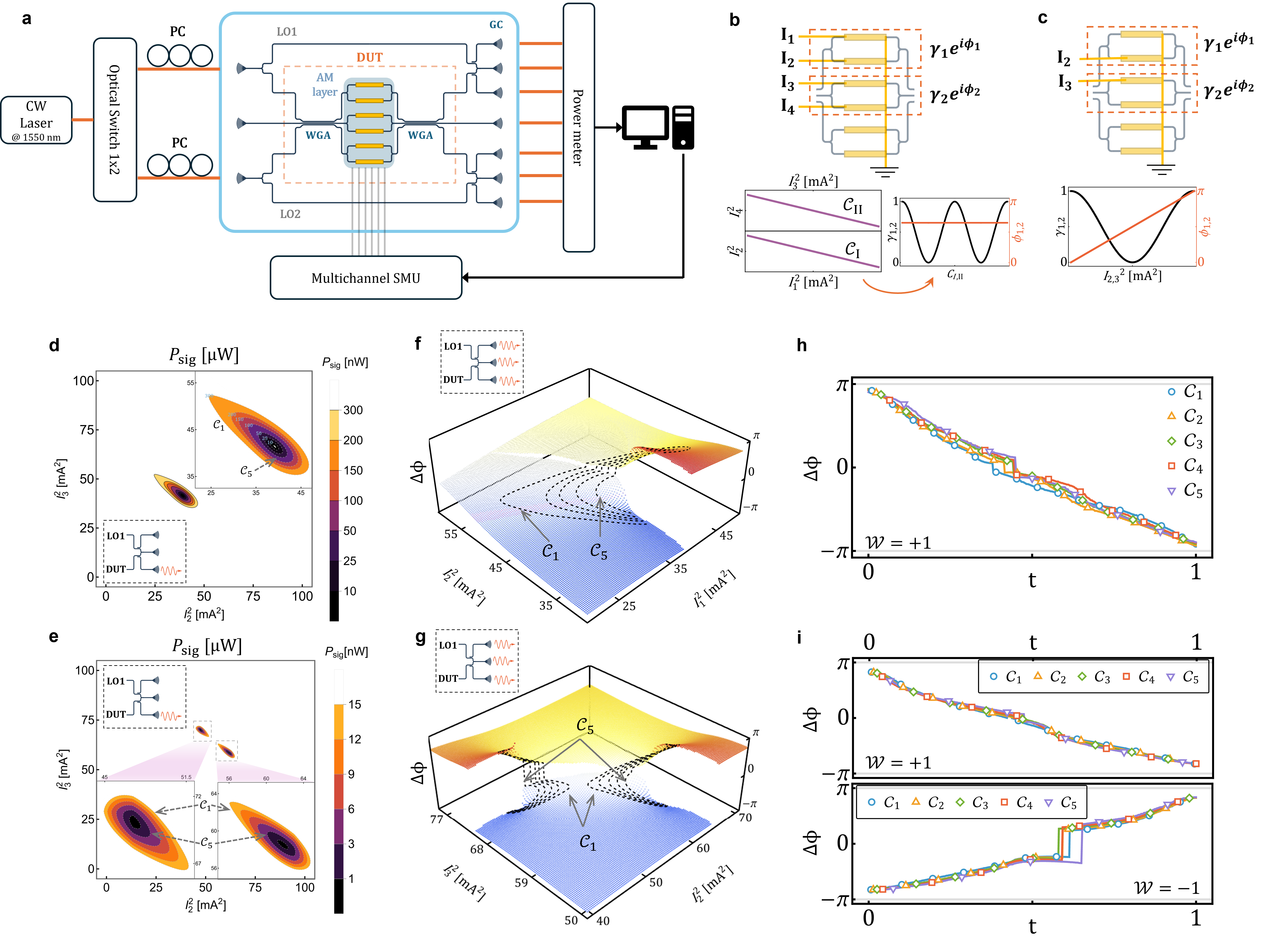}
    \caption{
    \textbf{Phase and power measurements.} 
    \textbf{a} Schematic experimental setup used for characterization. A 1550 nm CW laser, routed by a 1x2 switch and polarization controllers, excites the fundamental TE mode of the PIC. An external computer controls heater currents via an SMU and records the seven PIC outputs measured by an optical power meter. \textbf{b-c} Illustration of the MZI layer configured for: (\textbf{b}) four-heater operation (steady-phase mode) and (\textbf{c}) two-heater operation (amplitude-dependent phase mode) for each AM stage. \textbf{d–g} Experimental results of the synthetic phase-control unit for the steady-phase mode (\textbf{d}, \textbf{f}) and amplitude-dependent phase mode (\textbf{e}, \textbf{g}). \textbf{d–e} Equal-power closed contours at the signal sampling port ($P_{\text{sig}}$) encircling their isolated zeros. Insets show zoomed views. \textbf{f–g} Retrieved phase, $\Delta\phi$, plotted near the transmission zeros along with the contours $\mathcal{C}_j$ from \textbf{d–e}. Dashed rectangles indicate measurement ports.
    \textbf{h-i} Parametric plots $\mathcal{C}_{j}(t):(0,1)\rightarrow(-\pi,\pi)$ around the closed contours for both the steady-phase (\textbf{h}) and amplitude-dependent (\textbf{i}) operations. The upper and lower panels in~\textbf{i} correspond to the contours encircling the right and left zeros of \textbf{g}, respectively. 
    }
    \label{fig:Fig3}
\end{figure}

The experimental setup implemented to characterize and train the device is sketched in Fig.~\ref{fig:Fig3}\textbf{a}. A continuous-wave, wavelength-tunable laser is set to 1550 nm, the target wavelength, and connected to a multichannel optical switch. This switch enables selective excitation of the input grating couplers via an external computer. An 8-channel fiber array links the multichannel optical switch and the grating couplers. We used polarization paddles to optimize the coupling efficiency of the laser light to the grating couplers. The electrical channels are mapped to a PCB and interfaced with a source meter to supply voltage and current to the micro-heater element. The optical output channels are recorded through a multiport power meter. Each measurement consists of selectively exciting one input waveguide at a time and simultaneously capturing power from all output waveguide arrays after updating the driving current of each phase shifter. Due to the limited number of heaters (at most four simultaneously), we operate the PIC at room temperature and do not require an external thermoelectric cooler (TEC).

To experimentally validate the existence of closed trajectories $\mathcal{C}$ surrounding an isolated zero, we perform a grid search on the active elements. Each modulator is implemented using MZIs that include phase shifters $\phi_{1}$ and $\phi_{2}$ in each arm. The transmission $t_{\textnormal{mzi}}$ of an MZI is well-known and performs the non-power-conserving transformation 
\begin{equation}
t_{\textnormal{mzi}}=e^{i\Delta_{+}}\cos\left( \Delta_{-} \right), \quad \Delta_{\pm}:=\frac{\phi_{1}\pm\phi_{2}}{2} .
\label{eq:mzi}
\end{equation}
That is, each amplitude-modulation level is accompanied by a phase delay, a common side effect observed in other AMs, such as GST-based PCMs~\cite{yamada1991rapid}. In contrast, AM elements with a stable phase have been documented in the literature ~\cite{hatayama2024phase}. Given the tunability of the MZIs, we can apply the appropriate currents to the two heaters in each MZI to emulate both types of AMs, as described in Eq.~\eqref{eq:mzi}. Figures ~\ref{fig:Fig3}\textbf{b} and ~\textbf{c} illustrate the operation of each MZI. In Fig.~\ref{fig:Fig3}\textbf{b}, the parameter $\Delta_{+}$ is kept constant by operating both heaters in each MZI (four heaters in total). In contrast, in Fig.~\ref{fig:Fig3}\textbf{c}, only one heater per MZI is operated, resulting in a phase delay at each modulation step.

The device is operated by exciting the upper input port while measuring the power across all seven output ports. A first characterization is performed by keeping the phase in each MZI steady. To this, we sweep the currents $I_{1,2,3,4}$ across the four heaters on the top two MZIs so that $I_{1}^{2}\propto I_{2}^{2}$ and $I_{4}^{2}\propto I_{3}^{2}$, as illustrated in Fig.~\ref{fig:Fig3}\textbf{b}. The current square is used because the phase shift induced by each microheater is ideally linear with respect to the power, which is in turn proportional to the square of the current. The proportionality factor determines the overall steady-state phase of the MZI. 

Since we have only two independent currents, we plot in Fig.~\eqref{fig:Fig3}\textbf{d} the constant-power contours, $\mathcal{C}_{j}$, measured at the signal port on the sampling stage, which range from 50 nW to 300 nW. An additional contour of 10 nW is sampled, which is the smallest contour close to the transmission zero that can be characterized. The sampled contours follow elliptic-like trajectories, as predicted by the theoretical model. See Methods Section for details. 

A second characterization is done by operating only two of the six available heaters: the lower heater on the upper MZI ($I_{2}$) and the upper heater on the middle MZI ($I_{3}$). Here, we have performed a current sweep on both active heaters, varying their currents independently over the (0, 10.5) mA interval. Using the signal port on the sampling stage, we measured the resulting output power, $P_{\textnormal{sig}}$, and identified the combinations of heater currents that produce a constant signal power. These constant-power trajectories, $\mathcal{C}_{j}(I_{2}^{2},I_{3}^{2})$ are illustrated in Fig.~\ref{fig:Fig3}\textbf{e}. We have plotted five contours ($j\in\{1,\ldots,5\}$) spanning from 1.0 $\upmu$W to 5.5 $\upmu$W. An additional trajectory at 0.2 $\upmu$W (the black contour) was characterized to demonstrate the existence of an isolated zero. 

The power measurements swept across the contours $\mathcal{C}_{j}$ encircle their respective transmission zeros, confirming the protected topological winding number $\vert\mathcal{W}\vert=1$. 
To analyze the power in the interferometer port, $P_{\textnormal{int}}$, each closed contour $\mathcal{C}_{j}$ is parameterized by $t\in(0,1)$, with $t=0$ set to the $A$ marker on each contour. To extract the phase difference $\Delta\phi$, we use the power measurements from the LO1, interference, and signal ports, as indicated in the insets of Figs.~\ref{fig:Fig3}\textbf{f}-\textbf{g}. See Supplementary Information~\textcolor{red}{S3} for details about the phase retrieval process. The results are summarized in Figs.~\ref{fig:Fig3}\textbf{f}-\textbf{g} for MZIs operated in steady phase and non-steady phase, respectively. 

In the steady-phase case, we observe a single transmission zero surrounded by closed contours. The reconstructed phase response in Fig.~\ref{fig:Fig3}\textbf{f} exhibits the expected phase singularity gap associated with this zero. Furthermore, Fig.~\ref{fig:Fig3}\textbf{h} shows a parametric plot of the phase contours $\mathcal{C}_{i}$, where the full $2\pi$ phase range span is present across all the contours.

Conversely, in the amplitude-dependent case, two distinct transmission zeros emerge. We select contours that encircle each zero independently. This duality is a direct consequence of the nonlinear nature of the amplitude modulation; specifically, the modulation parameter induces a parasitic phase shift (see Supplementary Material \textcolor{red}{S2} for a detailed explanation). As illustrated in Fig.~\ref{fig:Fig3}\textbf{g}, a phase gap exists around each zero. This configuration provides the flexibility to operate around either transmission zero while still achieving a full phase rotation. In turn, Fig.~\ref{fig:Fig3}\textbf{i} illustrates the corresponding parametric plot of the phase span around each contour. The phase spans from $\pi\rightarrow - \pi$ and $- \pi\rightarrow \pi$ around the right and left zeros, respectively. This illustrates that the winding numbers around each zero have opposite signs ($\mathcal{W}=\pm 1$). 

\section*{Conclusion}
We have introduced a synthetic phase-control unit that operates using AMs as active components exclusively. By leveraging tunable losses and topologically protected quantities, our design replicates the functionality of an ideal phase shifter, enabling a complete $2\pi$ phase rotation at constant amplitude or steady phase with variable amplitude. This lossy approach addresses integration challenges, as AMs are typically smaller and faster than conventional thermo-optic phase shifters. 

As proof of concept and to showcase the versatility of the proposed unit, we have used MZIs as the AM units. This enables emulation of different AMs, with modulation performed using either a steady-phase or a variable-phase at each modulation level. Both functionalities can be achieved using MZIs, and in each case, the proposed phase-control unit can effectively perform its intended tasks. This eliminates the concern that phase control might be an artifact stemming from the amplitude-dependent phase characteristics of certain AMs. The experimentally characterized trajectories exhibited an elliptical-like shape, closely resembling the predicted conic sections from the theoretical model.

Beyond the previous operations, our experimental results highlight the dual functionality of the phase-control unit, which can transition between a constant-amplitude phase shifter and a constant-phase amplitude modulator by properly operating the amplitude modulators. This capability has recently been investigated in unconventional PCMs~\cite{hatayama2024phase}. We have also identified two singularities with opposite winding numbers ($\mathcal{W}=\pm 1$). Although a larger contour could be defined to encircle both singularities, the opposing charges would cancel out, resulting in a null winding number. We specifically exclude this case, as a non-zero topological charge is a strict requirement for ensuring the full $2\pi$ phase sweep necessary for robust control.

The topologically protected nature of the design provides robustness against fabrication defects, and its principles can be adapted to various AM technologies, opening new avenues for high-speed and flexible photonic systems. The primary application is the potential for programmable photonic linear transforms to enable optical information processing, providing an alternative, scalable mechanism for high-speed optical computing units. Moreover, the phase-control unit features a simple design that facilitates fabrication across diverse material platforms and ensures compatibility with open-access foundries, making the entire approach highly suitable for large-scale implementation. These results were experimentally validated using a proof-of-concept device based on a silicon-on-insulator platform and MZIs for amplitude modulation. However, it is worth noting that any AM solution, such as PCMs, doped-silicon modulators, or germanium-silicon modulators, can be used instead of the MZI. 

\section*{Methods}

\subsection*{Passive layer design}
The unitary matrix $F$ encapsulates the transmission characteristics of the waveguide array and the optical path differences in the fan-out region. See Fig.\textcolor{red}{S1}. The WGA is thus defined by the product
$$
F = \Phi_{\textnormal{fan-out}}^{T} e^{-i L H} \Phi_{\textnormal{fan-out}} ,
$$
where $\Phi_{\textnormal{fan-out}}=\textnormal{diag}\left(e^{i \phi_{1}},e^{i \phi_2},e^{i \phi_{3}}\right)$ accounts for the phases $\theta_{i}$ accumulated due to unbalanced paths at the WGA output. Although non-diagonal elements could theoretically model crosstalk, these effects are mitigated in this design and subsequently neglected. The term $\Phi_{\textnormal{fan-out}}^{T}$ arises from the reciprocity theorem, applicable due to the absence of magnetic materials. In turn, the WGA utilizes identical waveguides with homogeneous gaps, described by the tight-binding Hamiltonian $H_{n,m}=\frac{\widetilde{\kappa}}{\sqrt{2}}(\delta_{n+1,m}+\delta_{n-1,m})$, with $\widetilde{\kappa}$ a scaling factor derived from mode analysis and $L$ is the coupling length. To ensure maximum spatial-mode mixing, equivalent to a discrete fractional Fourier transform~\cite{Wei16,zelaya2024integrated}, the design parameters are fixed to satisfy $\widetilde{\kappa}L=\pi/2$.

\subsubsection*{On the existence of closed trajectories}
To showcase the existence of the transmission zeros and the corresponding closed trajectories around them, we consider the theoretical model in Eq.~\eqref{eq:DUT} with ideal AMs for illustration purposes. In the Supplementary Material \textcolor{red}{S1}, we show that at most one transmission zero exists in the $\gamma_{1}-\gamma_{2}$ plane, and the transmission at the output of the phase-control unit becomes
\begin{equation}
T=\frac{\gamma_{1}^{2}}{16}-\frac{\cos(\phi_1-\phi_2)}{4}\gamma_{1}\gamma_{2}+\frac{\gamma_{2}^{2}}{4}+\frac{\cos(\phi_{1})}{8}\gamma_{1}-\frac{\cos\phi_{2}}{4}\gamma_{2}+\frac{1}{16} ,
\label{eq:conic}
\end{equation}
where the phases $\phi_{1,2,3}$ are fixed and produced due to the optical path deviation at the fan-out of the waveguide array (see Fig. \textcolor{red}{S1}). Without loss of generality, we have fixed $\phi_{3}=0$, as we can always factor out a global phase. Eq.~\eqref{eq:conic} reveals that, for a fixed transmission efficiency $T$,  \textit{conic section} is spanned in the $\gamma_{1}-\gamma_{2}$ plane, parameterized by the optical path deviations produced by the phases $\phi_{1,2}$. 

Because only an elliptical path results in a closed loop, we need to determine if our conic section satisfies this condition. This can be determined from Eq.~\eqref{eq:conic} and the conic discriminant $\Delta=b^2-ac$ for a general conic of the form $ax^{2}+2bxy+cy^{2}+2dx+2ey+f=0$. This results in $\Delta=\sin^{2}(\phi_{1}-\phi_{2})/8$, a positive quantity for any optical phase deviation $\phi_{1}$. Thus, the existence of simple closed trajectories for our phase-control unit is always ensured, regardless of optical path deviations produced by the fan-out of the waveguide arrays. This explains the elliptic-like shape observed in the experimental characterization of the contours shown in Fig.~\ref{fig:Fig3}\textbf{d}.



\backmatter

\bmhead{Supplementary information}
The online version contains supplementary material.

\bmhead{Acknowledgements}
This project is supported by the U.S. Air Force Office of Scientific Research (AFOSR) Award\# FA9550-25-1-0200, and the Defense University Research Instrumentation Program (DURIP) Award\# FA9550-23-1-0539.

\bmhead{Data availability} 
Data underlying the results presented in this paper are not publicly available at this time but may be obtained from the authors upon reasonable request.

\bmhead{Competing interests}
The authors declare no competing interests.

\bmhead{Disclosures}
Patent pending.


\bibliography{Non_Hermitian_PS} 





\end{document}